\documentclass[ aip,
 amsmath,amssymb,
 graphicx,
 reprint,]{revtex4-1}

\usepackage[T1]{fontenc}
\usepackage[english]{babel}\usepackage{csquotes}\usepackage[activate={true,nocompatibility},tracking=true]{microtype}\usepackage{hyphenat}\usepackage{siunitx}
\usepackage{placeins}
\usepackage[begintext=``, endtext='']{quoting}
\SetBlockEnvironment{quoting}
\SetBlockThreshold{1}
\usepackage{hyperref}
\usepackage{subcaption}

\usepackage{amsthm}
\usepackage{amsmath}
\usepackage{amsfonts}
\usepackage{amssymb}
\usepackage{amscd}
\usepackage{mathtools}
\usepackage[mathscr]{eucal}
\usepackage{calligra}
\DeclareMathAlphabet{\mathcalligra}{T1}{calligra}{m}{n}
\DeclareFontShape{T1}{calligra}{m}{n}{<->s*[1.1]callig15}{}
\usepackage{bm}
\usepackage[nice]{nicefrac}
\usepackage{dsfont}

\usepackage{tikz}
\usepackage{pgf}
\usepackage{pgfplots}
\usepackage{pgfplotstable}
\usepackage{shellesc}
\pgfplotsset{compat=newest}
\usetikzlibrary{spy}
\usetikzlibrary{shapes}
\usetikzlibrary{backgrounds}
\usetikzlibrary{shadows}
\usetikzlibrary{matrix}
\usetikzlibrary{external}
\usetikzlibrary{positioning}
\usetikzlibrary{plotmarks}
\usepgfplotslibrary{fillbetween}
\usepackage{natbib}
\usepackage{comment}
\usepackage{graphicx}
\usepackage[colorinlistoftodos,prependcaption,textsize=small]{todonotes}
\usepackage[capitalize]{cleveref}

\crefname{equation}{}{}
\crefname{figure}{Figure}{Figures}
\usepackage{csvsimple}
\usepackage{booktabs}
\usepackage{environ}
\usepackage{adjustbox}
\usepackage{relsize}
\usepackage{afterpage}
\usepackage{booktabs}
\usepackage{arydshln}

\newcommand{\RomanNumeralCaps}[1]
\linenumbers

\let\min\relax \DeclareMathOperator*\min{\vphantom{p}min}

\let\tilde\widetilde
\let\hat\widehat

\DeclareMathOperator{\bcdot}{\bm{\cdot}}

\newcommand{\norm}[2][]{\left\|#2\right\|_{#1}}

\newcommand{\dom}{\mathcal{D}}
\newcommand{\loss}{\operatorname{MSE}}
\newcommand{\reg}{\operatorname{Reg}}
\newcommand{\pen}{\operatorname{Pen}}

\newcommand{\R}{\mathbb{R}}       \newcommand{\dd}{\,\mathrm{d}}

\newcommand{\bxi}{\bm{\xi}}

\newcommand{\bma}{\bm{a}}

\newcommand{\bmk}{\bm{k}}

\newcommand{\bmr}{\bm{r}}

\newcommand{\bmx}{\bm{x}}
\newcommand{\bmy}{\bm{y}}

\newcommand{\bmD}{\bm{D}}

\newcommand{\bmG}{\bm{G}}

\newcommand{\bmW}{\bm{W}}

\newcommand{\bmZ}{\bm{Z}}

\newcommand{\bsfD}{\bm{\mathsf{D}}}

\newcommand{\bfe}{\mathbf{e}}

\newcommand{\bfu}{\mathbf{u}}

\newcommand{\bfU}{\mathbf{U}}

\newcommand{\bfZ}{\mathbf{Z}}

\newcolumntype{?}{!{\vrule width 1.2pt}}

\makeatletter
\newsavebox{\measure@tikzpicture}
\NewEnviron{scaletikzpicturetowidth}[1]{	\tikzifexternalizingnext{		\def\tikz@width{#1}		\begin{lrbox}{\measure@tikzpicture}			\tikzset{external/export next=false,external/optimize=false}			\BODY
		\end{lrbox}		\pgfmathparse{#1/\wd\measure@tikzpicture}				\BODY
	}{		\BODY
	}}
\makeatother

\definecolor{color0}{rgb}{0.7843, 0.7843, 0.7843}
\definecolor{color1}{rgb}{0, 0.4470, 0.7410}
\definecolor{color2}{rgb}{0.8500, 0.3250, 0.0980}
\definecolor{color3}{rgb}{0.9290, 0.6940, 0.1250}
\definecolor{color4}{rgb}{0.7060, 0.3840, 0.7650}
\definecolor{color5}{rgb}{0.4660, 0.6740, 0.1880}
\definecolor{color6}{rgb}{0.3010, 0.7450, 0.9330}
\definecolor{color7}{rgb}{0.6350, 0.0780, 0.1840}
\definecolor{color8}{rgb}{0.0, 0.4078, 0.3412}

\pgfplotsset{
  log x ticks with fixed point/.style={
      xticklabel={
        \pgfkeys{/pgf/fpu=true}
        \pgfmathparse{exp(\tick)}        \pgfmathprintnumber[fixed relative, precision=3]{\pgfmathresult}
        \pgfkeys{/pgf/fpu=false}
      }
  },
  log y ticks with fixed point/.style={
      yticklabel={
        \pgfkeys{/pgf/fpu=true}
        \pgfmathparse{exp(\tick)}        \pgfmathprintnumber[fixed relative, precision=3]{\pgfmathresult}
        \pgfkeys{/pgf/fpu=false}
      }
  }
}

\tikzset{
  ashadow/.style={opacity=.25, shadow xshift=0.07, shadow yshift=-0.07},
}

\definecolor{CustomGreen}{RGB}{65,169,50}

\renewcommand{\vec}[1]{\bm{#1}}

\newcommand{\abs}[1]{\left| #1 \right|}

\newcommand{\ii}{\operatorname{i}}
\newcommand{\ee}{\operatorname{e}}

\newcommand{\LengthScale}{L}
\newcommand{\TimeScale}{T}
\newcommand{\Magnitude}{C}

 \renewcommand{\d}{\operatorname{d}}

\pgfplotsset{
    legend image with text/.style={
        legend image code/.code={            \node[anchor=center] at (0.3cm,0cm) {#1};
        }
    },
}

\makeatletter
\tikzset{
    scale plot marks/.is choice,
    scale plot marks/false/.code={
        \def\pgfuseplotmark##1{\pgftransformresetnontranslations\csname pgf@plot@mark@##1\endcsname}
    },
    scale plot marks/true/.style={},
    scale plot marks/.default=true
}
\makeatother
\graphicspath{{./Figures/}}
\definecolor{CeruleanRef}{RGB}{12,127,172}
\hypersetup{colorlinks=true,allcolors=CeruleanRef}
\hypersetup{pdfstartview=FitB,pdfpagemode=UseNone}

\makeatletter
\def\@email#1#2{ \endgroup
 \patchcmd{\titleblock@produce}
  {\frontmatter@RRAPformat}
  {\frontmatter@RRAPformat{\produce@RRAP{*#1\href{mailto:#2}{#2}}}\frontmatter@RRAPformat}
  {}{}
}\makeatother
\begin{document}

\title{Learning the structure of wind: A data-driven nonlocal turbulence model for the atmospheric boundary layer}

\author{B. Keith}
  \protect\thanks{These two authors contributed equally.}
 \email[electronic mail and correspondence:~]{keith10@llnl.gov}
 \affiliation{ 
Center for Applied Scientific Computing, Lawrence Livermore National Laboratory, U.S.A.
}
\author{U. Khristenko   } \thanks{These two authors contributed equally.}
\author{B. Wohlmuth}
\affiliation{ 
Department of Mathematics, Technical University of Munich, Germany
}

\date{\today}

\begin{abstract}
	We develop a novel data-driven approach to modeling the atmospheric boundary layer.
	This approach leads to a nonlocal, anisotropic synthetic turbulence model which we refer to as the deep rapid distortion (DRD) model.
		Our approach relies on an operator regression problem which characterizes the best fitting candidate in a general family of nonlocal covariance kernels parameterized in part by a neural network.
	This family of covariance kernels is expressed in Fourier space and is obtained from approximate solutions to the Navier--Stokes equations at very high Reynolds numbers.
	Each member of the family incorporates important physical properties such as mass conservation and a realistic energy cascade.
	The DRD model can be calibrated with noisy data from field experiments.
	After calibration, the model can be used to generate synthetic turbulent velocity fields.
	To this end, we provide a new numerical method based on domain decomposition which delivers scalable, memory-efficient turbulence generation with the DRD model as well as others.
	We demonstrate the robustness of our approach with both filtered and noisy data coming from the 1968 Air Force Cambridge Research Laboratory Kansas experiments.
	Using this data, we witness exceptional accuracy with the DRD model, especially when compared to the International Electrotechnical Commission standard.
								\end{abstract}

\maketitle

\section{Introduction} \label{sec:introduction}

The purpose of this paper is to present a new turbulence model for the atmospheric boundary layer (ABL).
The model is derived in part using a classical technique to study high $Re$ flows, called rapid distortion theory (RDT).\cite{townsend1980structure}
We refer to this model throughout the text as the deep rapid distortion (DRD) model.

The ABL is the bottom layer of the troposphere, extending from the ground usually to the first few hundred meters above Earth's surface.
In the ABL, the Reynolds number is very large\cite{wyngaard2010turbulence} ($Re\sim$\num{e6}--\num{e8}), which hampers many high-fidelity and mid-fidelity modeling techniques.
Direct numerical simulations (DNSs) have long been ruled out, with even the most massive simulations of canonical flows in recent years falling short of the $Re\sim $ \num{e5} threshold.\cite{lee_moser_2015,li2008public}
Wall-modeled large eddy simulations (LESs) of atmospheric flows make up for the DNS gap by relying on well-calibrated subgrid-scale (SGS) models.\cite{stoll2020large}
These coarse-mesh models deliver the correct mean energy transfer to small scales, but often fail to accurately capture the true SGS dynamics.\cite{meneveau2000scale,moser2021statistical}
In wall-resolved LES, on the other hand, the promise of making up for the DNS gap has only been partially realized.\cite{lohner2019towards}
This is largely due to the fact that many industrial-scale fine-mesh LESs at $Re\sim$\num{e6} presently require no less than \num{3}--\num{4} weeks to complete, regardless of the size of the machine.\cite{lohner2019towards}
Such long simulation times are prohibitive in many engineering applications.

To only stymie matters further, many ``big data'' methods, which have been successfully used to model lower $Re$ flows, require rich and well-structured DNS or LES data.\cite{lohner2019towards}
Because such data sets are mostly unattainable for ABL flows, alternative modeling approaches need to be considered.

One successful approach to modeling at very large $Re$ is to adopt a statistical characterization of the flow.\cite{kraichnan1970diffusion,mann1994spatial,Veers1988,rinker2018pyconturb}
Such approaches may rely in part on laboratory experiments or field measurements.
This provides the additional benefit of incorporating latent information about the environment that cannot be extracted from any simulation alone.
Such models are also important for coupling meso- and micro-scale models and providing inflow data for high-resolution large eddy simulations.\cite{lee1992simulation,JARRIN2006585,tabor2010inlet,talbot2012nested,munoz2015stochastic,wu2017inflow,haupt2020mesoscale,zhong2021implementation}

Turbulence has long been analyzed from a statistical point of view.\cite{taylor1935statistical}
We seek to continue that enterprise with the assistance of new machine learning (ML) technology.
Instead of training an operator which emulates flow features from simulation data sets, we choose to learn the spectral structure of the flow field from (possibly noisy) experimental measurements.
Our approach enforces an explicit structure on the model which is derived directly from the underlying incompressible Navier--Stokes equations.
Contrary to ``black box'' scientific ML techniques, our approach delivers a human-understandable model which can be analyzed after training to infer subtle physical characteristics.
The class of models we work with, spectral tensor models,\cite{townsend1980structure,Maxey1982,Hunt1984turbulence,mann1994spatial,mann2000spectral,sanderse2011review,chougule2017modeling,Chougule2018} are not new to fluid dynamics, but are herein reimagined in a new scientific ML context.

Our task is to model homogeneous neutral ABL turbulence as a divergence-free random velocity field.
In many engineering applications, only second-moment statistical data are available.\cite{simiu2019wind}
Fortunately, this is enough to characterize many of the most important physical quantities in the flow.\cite{pope2001turbulent}
In such cases, it is appropriate to assume that the field is Gaussian.
After centering the mean, the problem then reduces to isolating the appropriate covariance operator.
Because of homogeneity, this can be done by focusing on the stationary spectral velocity tensor.\cite{frehlich2001simulation}
Extensions of our approach to non-Gaussian and inhomogeneous fields can be made.\cite{bolin2014spatial,wallin2015geostatistical,keith2020fractional}
Our approach can also be extended to include thermal coupling effects.\cite{Chougule2018}
We mention some of these possibilities, but they are not part of this work.

Unlike some model inference techniques of contemporary interest,\cite{anandkumar2020neural,lu2021learning} we only seek to learn a locally bounded kernel.
This decision leads to a much easier ML problem than seeking out the full integro-differential operator, which is typically unbounded.\cite{ciarlet2013linear}
Several model inference strategies focus on integral kernels\cite{feliu2020meta,gin2020deepgreen,boulle2021data} and others rely on Fourier transforms to facilitate the learning process.\cite{li2021fourier,PATEL2021113500}
Our approach, however, stands apart through its unique inductive biases and focus on random fields instead of PDEs.

We begin with a solution of the linearized Navier--Stokes equations in a uniform mean shear flow based on RDT.\cite{townsend1980structure}
This classical solution provides a time-dependent parameterization of the spectral velocity tensor.\cite{Maxey1982}
Stationarity is then imposed by substituting a wavevector-dependent eddy lifetime for the temporal parameter.
Numerous eddy lifetime models have appeared in the classical literature,\cite{comte1971simple,lesieur1987turbulence,mann1994spatial} however, we note that this form of substitution was first introduced by Derbyshire and Hunt\cite{derbyshire1992structure} and later popularized by Mann.\cite{mann1994spatial,mann1998wind}

Like Mann, our approach largely centers on selecting the eddy lifetime.
Also like Mann, we choose to calibrate the resulting model with one-point spectra data from the 1968 Air Force Cambridge Research Laboratory Kansas experiments.
In comparing our results, we witness the exceptional accuracy of the DRD model.
This is due, in part, to our use of a specially-designed neural network architecture to parameterize the eddy lifetime.

The DRD model may be used as a substitute for the International Electrotechnical Commission (IEC) 61400-1 standard,\cite{tc882005iec,mann1998wind} which is widely used in wind turbine design\cite{sanderse2011review,keck2014synthetic} and other wind engineering applications.\cite{gawronski2007servo,gawronski2007control,Michalski2011,Andre2015,MICHALSKI201596}
It also provides an alternative which bypasses the computational limitations of other data-driven models for turbulence generation that have been applied in lower $Re$ environments and rely on large datasets.\cite{chertkov2018deep,bode2018towards,fukami2019synthetic,kim2020deep,subramaniam2020turbulence}

In order to distinguish our work from many others in the rapidly developing subdiscipline of ML for fluid mechanics,\cite{brunton2020machine} we itemize the main contributions of this work.
Our main contributions include:
\begin{itemize}
    \item
    We introduce a nonlocal, anisotropic, data-driven turbulence model for wind in the ABL (the DRD model).
    The DRD model is more accurate than a simpler spectral model standardized by IEC,\cite{tc882005iec} yet it is no more expensive to use.
                    \item
    Unlike ``black box'' ML approaches to model discovery and synthetic turbulence generation, our approach is based on basic physical principles and derived directly from the Navier--Stokes equations.
    Furthermore, the model we arrive at is human-interpretable, since all of its information is explicitly encoded in a kernel function; i.e, the spectral tensor.
    \item
    We present a simple, efficient, and mathematically justified numerical method for generating synthetic turbulence.
    This contribution is two-fold, as the method can be applied to the DRD turbulence model just as easily as it can be applied to other spectral models used in practice.
    \item
    We release our code,\cite{PythonCode} implemented in the Python programming language and leveraging the PyTorch\cite{NEURIPS2019_9015} software library, with the intention of adoption of the proposed methods in the wider community.
\end{itemize}

The remainder of the paper is structured as follows.
In the next section, we introduce fundamental notation and concepts related to the ABL.
In~\cref{sec:Model}, we present the DRD model.
In~\cref{sec:Calibration}, we describe the model calibration process.
In~\cref{sec:implementation_details}, we introduce a novel domain decomposition technique for turbulence generation.
Finally, in~\cref{sec:conclusion}, we close with a brief overview of results.

\section{Wind in the atmospheric boundary layer} \label{sec:atmospheric_boundary_layer_flow}

The ABL is characterized by constant shear stress in the vertical direction $\bfe_3$ and is generally recognized to be \emph{neutrally stable} at high wind speeds.\cite{kaimal1994atmospheric}
That is, the buoyancy forces due to temperature gradients may be assumed to be negligible in comparison to surface-driven friction forces.
These physical assumputions are important to this work.
Relying on them, we will derive a statistical model which characterizes the ABL as a mass-conserving random field $\bfU$ with homogeneous turbulent fluctuations $\bfu$.

In this work, we denote the atmospheric turbulent velocity field as $\bfU(\bmx) = (U_1(\bmx),U_2(\bmx),U_3(\bmx))$, where $\bmx = (x_1, x_2, x_3) = (x, y, z)$ denotes standard Cartesian coordinates with the $x$-axis indicating the direction of the mean wind field and $z$ indicating the height above ground.
The turbulent fluctuations around the mean wind field are denoted $\bfu = \bfU - \langle \bfU\rangle = (u_1, u_2, u_3)$, with $\langle \,\cdot\, \rangle$ denoting the ensemble average.
In this section, we review several common modeling assumptions for both the mean velocity profile $U(z)$ and the turbulent fluctuations $\bfu(\bmx)$ in the ABL.

\subsection{Mean profile} \label{sub:mean_profile}

Ground friction is dominated by pressure drag, which is a force generated by pressure differences near the surface and caused by wind flowing across surface obstacles.
Depending on the local terrain, the variety of obstacles affecting ground friction can change vastly.
For instance, consider that different friction forces will arise from flow across grass, forests, open water, or urban canopies.
In this study, we strongly simplify the many effects of ground friction and assume that the mean wind field may be represented by a \emph{uniform shear} flow in the $x$-direction, $\langle \bfU(\bmx)\rangle = \langle U_1(z)\rangle\,\bfe_1$, where
\begin{equation}
\label{eq:MeanProfile}
	\langle U_1(z)\rangle = A + B z,
	\qquad
	A,B \geq 0
	.
\end{equation}

In our approach, the ABL turbulent fluctuations $\bfu(\bmx)$ are modeled as though they have been distorted by the shearing mean profile~\cref{eq:MeanProfile}.
This approximation will only be appropriate to first order, locally in $z$, and will degrade as the length scale of the eddies grows.
Nevertheless, such simplified models are well-established in the literature; see, e.g., Ref.~\onlinecite[p.~145]{mann1994spatial}.

In most terrain categories, \cite{jcss2001probabilistic} a logarithmic profile such as 
\begin{equation}
\label{eq:MeanProfileLog}
	\langle U_1(z)\rangle = \frac{u_\ast}{\kappa}\ln\bigg(\frac{z}{z_0}+1\bigg)
	,
		\end{equation}
$u_\ast,z_0>0,~\kappa \approx 0.41$, is quite appropriate for larger length scales, yet is unnecessary for our present purposes.
 In this expression, $z_0$ is generally referred to as the roughness height, $u_\ast$ is the friction velocity, and $\kappa$ is the von K\'{a}rm\'{a}n constant.
In~\cref{sec:Model}, we will show that~\cref{eq:MeanProfile} leads to a family of models with free parameters which can be calibrated with experimental data to partially overcome its discrepancy from~\cref{eq:MeanProfileLog} and other forthcoming simplifying assumptions.

\subsection{Spectra} \label{sub:spectra}

We have set out to base our model only on the second-moment statistical structure of the ABL.
Holding true to this aim, we will neglect higher moment modeling errors and assume that the fully developed turbulent velocity field $\bfu$ is Gaussian.
In turn, we focus on the two-point correlation tensor
\begin{equation*}
	R_{ij}(\bmr,\bmx)
	=
	\langle u_i(\bmx)u_j(\bmx+\bmr)\rangle
		.
\label{eq:CovarianceTensor}
\end{equation*}

In spatially homogeneous flows, we have that $R_{ij} (\bmr,\bmx) = R_{ij}(\bmr)$, and so it is convenient to transform the correlation tensor to Fourier space.
This transformation defines the velocity-spectrum tensor,
\begin{equation*}
\label{eq:StationarySpectralTensor}
\Phi_{ij}(\bmk)
=
\frac{1}{(2\pi)^3}\int_{\R^3} \ee^{-\ii\bmk\bcdot\bmr}R_{ij}(\bmr)\dd \bmr
,
\end{equation*}
where $\bmk = (k_1,k_2,k_3)$ is the wavevector.
Due to the cross-correlation theorem, the velocity-spectrum tensor can also be written
\begin{equation}
\label{eq:StationarySpectralTensor2}
	\Phi_{ij}(\bmk)
	=
	\langle \overline{\hat{u}}_i(\bmk)\hat{u}_j(\bmk)\rangle
	,
\end{equation}
where $\hat{\bfu} = (\hat{u}_1,\hat{u}_2,\hat{u}_3)$ denotes the Fourier transform of $\bfu$.

A standard form of the spectral tensor used in isotropic, stationary, homogeneous turbulence models is
\begin{equation}
\Phi_{ij}^{\mathrm{VK}}(\bmk)
=
\frac{E(k)}{4\pi k^{2}}\,
\bigg(
	\delta_{ij} - \frac{k_ik_j}{k^2}
\bigg)
\,,
\label{eq:IsotropicSpectralTensor}
\end{equation}
where $k = |\bmk|$ is the wavenumber (the magnitude of the wavevector) and $E(k)$ is called the {energy spectrum function}.
A common empirical model for $E(k)$, dating back to von K\'arm\'an,\cite{von1948progress} is given by the expression
\begin{equation}
E(k)
=
c_0^2\, \varepsilon^{2/3}k^{-5/3}
\bigg(
\frac{k L}{(1 + (k L)^2)^{1/2}}
\bigg)^{17/3}
.
\label{eq:VKEnergySpectrum}
\end{equation}
Here, $\varepsilon$ is the viscous dissipation of the turbulent kinetic energy, $L$ is a length scale parameter, and $c_0^2\approx 1.7$ is an empirical constant.
The von K\'arm\'an model $\Phi^{\mathrm{VK}}(\bmk)$ encodes important nonlocal information about isotropic turbulence.
In fact, the energy spectrum~\cref{eq:VKEnergySpectrum} permits the convenient description of such turbulence as the solution of a (nonlocal) fractional diffusion equation.\cite{keith2020fractional}

It is typically not possible to directly measure a spectral tensor $\Phi_{ij}$ in a shearing flow.
Instead, one often collects the one-point spectra\cite{simiu2019wind}
\begin{equation*}
		F_{ij}(k_1)
	=
	\frac{1}{2\pi} \int_{-\infty}^\infty R_{ij} ((r_1,0,0) \ee^{-\ii k_1\cdot r_1}\! \dd r_1
	,
			\end{equation*}
with $i,j=1,2,3$.

Obviously, the one-point spectra cannot give a complete description of the turbulent wind field.
For this reason, it is necessary to construct physical models which can fit these experimental observations.
One of the issues in constructing these models is that turbulence in the atmospheric boundary layer is not spatially homogeneous.
Nevertheless, the common surface layer scaling assumption\cite{kaimal1972spectral} is that the length scales are proportional to $z$ and the variances are proportional to $u_\ast^2$.
For instance, after analyzing measurements taken above flat homogeneous terrain in Kansas, Kaimal et al.\cite{kaimal1972spectral} proposed taking
\begin{subequations}
\label{eq:Kaimal}
\begin{equation}
\label{eq:KaimalSu}
			\frac{k_1 F_{11}(k_1 z)}{u_\ast^2}
	=
	J_1(f)
	:=
	\frac{52.5 f}{(1 + 33f)^{5/3}}
	,
\end{equation}
\begin{equation}
\label{eq:KaimalSv}
			\frac{k_1 F_{22}(k_1 z)}{u_\ast^2}
	=
	J_2(f)
	:=
	\frac{8.5 f}{(1 + 9.5 f)^{5/3}}
	,
\end{equation}
\begin{equation}
\label{eq:KaimalSw}
			\frac{k_1 F_{33}(k_1 z)}{u_\ast^2}
	=
	J_3(f)
	:=
	\frac{1.05 f}{1 + 5.3f^{5/3}}
	,
\end{equation}
where $f = (2\pi)^{-1} k_1 z$, alongside $F_{12} = F_{23} = 0$ and
\begin{equation}
\label{eq:KaimalCross}
	-\frac{k_1 F_{13}(k_1 z)}{u_\ast^2}
	=
	J_4(f)
	:=
	\frac{7 f}{(1 + 9.6f)^{12/5}}
	.
\end{equation}
\end{subequations}
In \cref{sec:Calibration}, we will use these equations as filtered measurement data for benchmarking the DRD model.

\section{Model}
\label{sec:Model}

In this section, we derive a family of velocity spectrum models for the ABL.
After introducing more preliminary notation, we derive a time-dependent parameterization $\Phi(\bmk,\tau)$ from a linearized form of the Navier--Stokes equations.
We then enrich this parameterization with a general eddy lifetime model $\tau = \tau(\bmk)$ which includes physical symmetries and energy constraints.

\subsection{Gaussian hypothesis} \label{sub:preliminaries}

As with the two-point correlation tensor, it is convenient to consider the Fourier transform of the velocity field $\bfu$.
In such cases, we express the field in terms of a generalized Fourier--Stieltjes integral,
\begin{equation}
\bfu(\bmx)
=
\int_{\R^3}
\ee^{\ii\bmk\cdot\bmx}
\dd \bmZ(\bmk)
\,,
\label{eq:FourierStieltjes}
\end{equation}
where $\bmZ(\bmk)$ is a three-component signed measure on $\R^3$.\cite{lord2014introduction}

Let us consider three-dimensional additive white Gaussian noise~\cite{hida2013white,kuo2018white} in the physical and frequency domains, denoted $\bxi(\bmx)$ and $\hat{\bxi}(\bmk)$, respectively, such that
\begin{equation}\label{eq:ksi}
\bxi(\bmx)
=
\int_{\R^3} \ee^{\ii\bmk\cdot\bmx} \hat{\bxi}(\bmk) \dd \bmk
=
\int_{\R^3} \ee^{\ii\bmk\cdot\bmx} \dd \bmW(\bmk),
\end{equation}
where $\bmW(\bmk)$ is three-dimensional Brownian motion.
Our Gaussian assumption is that $\dd \bmZ(\bmk) = \bmG(\bmk) \dd \bmW(\bmk) = \bmG(\bmk)\, \hat{\bxi}(\bmk) \dd \bmk$, where $\bmG(\bmk)\, \bmG^\ast(\bmk) = \Phi(\bmk)$.

\subsection{Rapid distortion} \label{sub:rapid_distortion}

The rapid distortion equations~\cite[see, e.g.,][]{townsend1980structure,Hunt1990,pope2001turbulent} are a linearization of the Navier--Stokes equations in free space which holds when the turbulence-to-mean-shear time scale ratio is arbitrarily large.
To write them, we must first define the average total derivative of the turbulent fluctuations, namely
\begin{equation*}
\frac{\bar{D} u_i}{\bar{D} t}
=
\frac{\partial u_i}{\partial t}
+
\langle U_j\rangle\frac{\partial u_i}{\partial x_j}
\,.
\end{equation*}
With this definition in hand, the rapid distortion equations are
\begin{subequations}
\label{eq:RapidDistortionEquations}
\begin{align}
\label{eq:RapidDistortionEquations1}
\frac{\bar{D} u_i}{\bar{D} t}
&=
-u_i \frac{\partial \langle U_j\rangle}{\partial x_i} - \frac{1}{\rho}\frac{\partial p}{\partial x_i}
,
\\
\label{eq:RapidDistortionEquations2}
\frac{1}{\rho} \Delta p
&=
-2\frac{\partial \langle U_i\rangle}{\partial x_j}\frac{\partial u_j}{\partial x_i}
\,,
\end{align}
\end{subequations}
where $\rho$ and $p$ stand for the mass density and the hydrostatic pressure, respectively.

Some exact solutions of~\cref{eq:RapidDistortionEquations} can be found in the classical literature.\cite{townsend1980structure}
We are unaware of any closed-form solution with the mean profile in~\cref{eq:MeanProfileLog}.
However, with the uniform-shear mean profile~\cref{eq:MeanProfile}, the solution can be written in terms of an evolving wavevector $\bmk(t) = (k_1(t),k_2(t),k_3(t))$ and a non-dimensional time parameter $\tau = B t$.
For greater perspective, we begin with the general case where $\partial \langle U_i\rangle / \partial x_j$ is a constant tensor.

First, we define the rate of change of each frequency $\bmk(t)$ as follows:
\begin{equation}	
	{\!\dd k_i}/{\!\dd t} = -k_j{\partial \langle U_j\rangle}/{\partial x_i}.
\end{equation}
We then have the following Fourier representation of the average total derivative of~$\bfu$:
\begin{equation}
\begin{aligned}
\frac{\bar{D} u_i}{\bar{D} t}
&=
\int_{\R^3}
\ee^{\ii\bmk\cdot\bmx}\Bigg(
\bigg(\frac{\partial }{\partial t} + \frac{\dd k_j}{\dd t}\frac{\partial }{\partial k_j}\bigg) \dd Z_i(\bmk,t)
\Bigg)
\\
&=
\int_{\R^3}
\ee^{\ii\bmk\cdot\bmx}\Bigg(\frac{\bar{D} \dd Z_i(\bmk,t)}{\bar{D} t}\Bigg)
.
\end{aligned}
\end{equation}
With these expressions, the Fourier representation of~\cref{eq:RapidDistortionEquations} amounts to
\begin{equation}
\label{eq:RapidDistortionEquationsFourier}
\frac{\bar{D} \dd Z_j(\bmk,t)}{\bar{D} t}
=
\frac{\partial U_\ell}{\partial x_k}
\bigg(
2 \frac{k_j k_\ell}{k^2} - \delta_{j\ell}
\bigg)
\dd Z_k(\bmk,t)
\,,
\end{equation}
which must be accompanied by the initial state $\dd \bmZ(\bmk(0),0) = \dd \bmZ_0(\bmk(0))$.
From now on, we use the notation $\bmk_0 = \bmk(0) = (k_{10},k_{20},k_{30})$ to denote the initial wavevector.

In the scenario found in~\cref{eq:MeanProfile}, namely $\langle\bfU\rangle = (A + B z, 0, 0)$, the solution to~\cref{eq:RapidDistortionEquationsFourier} may be written as follows:
\begin{equation}
\label{eq:RDsolution}
\dd \bfZ(\bmk(t),t)
=
\bmD_\tau(\bmk)
\dd \bfZ(\bmk_0,0)
,
\end{equation}
where
\begin{equation*}
\bmD_\tau(\bmk)
=
\begin{bmatrix*}
1 & 0 & \zeta_1\\
0 & 1 & \zeta_2\\
0 & 0 & \zeta_3
\end{bmatrix*}
,
\qquad
\bmk = \begin{bmatrix}
1 & 0 & 0 \\
0 & 1 & 0 \\
-\tau & 0 & 1
\end{bmatrix}
\bmk_0
.
\end{equation*}
In the expression for $\bmD_\tau(\bmk)$, the non-dimensional coefficients $\zeta_i = \zeta_i(\bmk,\tau)$, $i=1,2,3$, are defined
\begin{equation*}
\zeta_1 = C_1 - C_2 k_2/k_1,
\quad
\zeta_2 = C_1 k_2/k_1 + C_2,
\quad
\zeta_3 = k_0^2/k^2,
\end{equation*}
where $k_0 = |\bmk_0|$ and
\begin{align*}
C_1 &= \frac{\tau k_1^2(k_0^2 -2k_{30}^2 + \tau k_1 k_{30})}{k^2(k_1^2+k_2^2)},
\\
C_2 &= \frac{k_2 k_0^2}{(k_1^2+k_2^2)^{3/2}}
\arctan x\biggr|_{k_3/\sqrt{k_1^2 + k_2^2}}^{k_{30}/\sqrt{k_1^2 + k_2^2}}
.
\end{align*}

All that is left is to set the initial state $\dd \bmZ(\bmk_0,0) = \dd \bmZ_0(\bmk_0)$.
We define the initial state to be a Gaussian random field,
\begin{equation}
\label{eq:RDGaussian}
	\dd \bmZ_0(\bmk_0)
	=
			\bmG_0(\bmk_0) \dd \bmW(\bmk_0),
\end{equation}
with $\bmG_0(\bmk_0)$ induced by the isotropic spectral tensor~\eqref{eq:IsotropicSpectralTensor} as follows,
\begin{equation*}
	\bmG_0(\bmk_0)\, \bmG_0^\ast(\bmk_0) = \Phi^{\mathrm{VK}}(\bmk_0)
	.
\end{equation*}

We have now uniquely defined $\Phi(\bmk,\tau)$.
First, following from~\cref{eq:ksi,eq:RDsolution,eq:RDGaussian}, we may write $\hat{\bfu}(\bmk) = \bsfD_\tau(\bmk)\, \bmG_0(\bmk_0)\, \hat{\bxi}(\bmk_0)$.
Second, by~\cref{eq:StationarySpectralTensor2} and the identity $\big\langle \overline{\hat{\xi}}_i \hat{\xi}_j\big\rangle = \delta_{ij}$, we have that
\begin{align*}
	\Phi(\bmk,\tau)
	&=
	\big\langle \hat{\bfu}(\bmk) \hat{\bfu}^\ast(\bmk)  \big\rangle
	\\
	&=
	\bsfD_\tau(\bmk)\, \bmG_0(\bmk_0) \big\langle \hat{\bxi}(\bmk_0)\ \hat{\bxi}^\ast(\bmk_0) \big\rangle\, \bmG_0^\ast(\bmk_0)\, \bsfD_\tau^\ast(\bmk)
	\\
	&=
	\bsfD_\tau(\bmk)\, \bmG_0(\bmk_0) \, \bmG_0^\ast(\bmk_0)\, \bsfD_\tau^\ast(\bmk)
	\\
	&=
	\bsfD_\tau(\bmk)\, \Phi^{\mathrm{VK}}(\bmk_0)\, \bsfD_\tau^\ast(\bmk)
	.
\end{align*}

\subsection{Eddy lifetime}

The spectral tensor $\Phi(\bmk,\tau)$ characterizes a time-dependent family of anisotropic covariance kernels which are stretched in the direction of the constant mean shear.
One flaw in this model is that physical eddies will break apart after a certain amount of distortion.

An important extension of the rapid distortion model involves replacing the distortion parameter~$\tau$ by a wavenumber-dependent \emph{eddy lifetime} function $\tau(\bmk)$.
The benefit of this substitution is two-fold: not only does it produce a stationary (time-\emph{independent}) spectral tensor $\Phi(\bmk,\tau(\bmk))$, but it provides a mechanism to recover missing physics which are neglected in the original modeling assumptions.

Various eddy lifetime models have been proposed in the literature;\cite{comte1971simple,lesieur1987turbulence,mann1994spatial} each of which involves restricting the eddy lifetime~$\tau$ to a radial function of the wavevector, $\tau = \tau(k)$.
For instance, a widely used radial function is presented in Ref.~\onlinecite{mann1994spatial} and results in a spectral tensor model which was subsequently standardized by the IEC.\cite{tc882005iec}
In this approach, the destruction of an eddy with size $k^{-1}$ is assumed to be mainly due to eddies of comparable or smaller size.
Thus, the square of the characteristic velocity of all influential eddies is given by $\int_{k}^{\infty}E(p)\dd p$.
By matching units, the eddy lifetime~$\tau$ may be proportional to a length scale divided by a velocity , e.g.,
\begin{equation}\label{key}
	\tau \propto k^{-1} \left[\int_{k}^{\infty}E(p)\dd p\right]^{-\frac{1}{2}}
	.
\end{equation}
After substituting~\cref{eq:VKEnergySpectrum}, this expression results in
\begin{equation}\label{eq:tau_Mann}
\tau^{\rm IEC}(k)
	= \frac{\TimeScale B^{-1} \, (kL)^{-\frac{2}{3}}}{\sqrt{_2F_1(1/3,17/6;4/3;-(kL)^{-2})}},
\end{equation}
where $_2F_1(a,b;c;x)$ denotes the hypergeometric function, and the time scale~$\TimeScale$ is a free parameter.
To facilitate an accurate comparison to this model in \Cref{sec:Calibration}, we point out that
\begin{equation*}
	\tau^{\rm IEC}(k) \propto
	\begin{cases}
		k^{-1} \quad &\text{for } k\to 0,\\
		k^{-3/2} \quad &\text{for } k\to \infty.\\
	\end{cases}
\end{equation*}

Substituting~\cref{eq:tau_Mann} into the rapid distortion spectral tensor, results in the Mann uniform shear model,
\begin{equation}
\label{eq:IECTensor}
	\Phi^{\rm IEC}(\bmk)
	=
	\Phi(\bmk,\tau^{\rm IEC}(|\bmk|))
	.
\end{equation}
Expanded formulas for $\Phi^{\rm IEC}(\bmk)$ can be found in Refs.~\onlinecite{mann1994spatial,tc882005iec}.

Given the anisotropic nature of shear flow, it is natural to expect that the lifetime of an eddy could depend on its initial spatial orientation.
However, it is also clear that this dependence should satisfy some basic physical conditions.
For instance, owing to translational symmetry of the flow in the transversal direction ($y$-axis), we expect $\tau(\vec{k})$ to be an even function of $k_2$.
We note that this restriction guarantees that the transversal components of the one-point cospectra, $F_{12}$ and $F_{23}$, vanish; cf.~\cref{eq:Kaimal}.

We now propose a method to discover complete wavevector eddy lifetime functions $\tau(\bmk)$ from measured data.
To the best of our knowledge, this is the first effort to accomplish this task.
Our approach involves writing $\tau(\bmk)$ as a feedforward neural network.

\subsection{Neural network model for the eddy lifetime}
\label{sec:tauNet}

Our goal is to look for $\tau$ in the form of a neural network with the wavevector~$\vec{k}$ as input.
Then, learning its parameters by fitting the one-point spectra data, we identify the function~$\tau(\vec{k})$ as well as the complete spectral tensor model.

Aside from physical symmetries, the eddy lifetime must satisfy certain asymptotic behavior in the limits $k\to 0$ and $k\to\infty$.
To this end, we first rewrite the eddy lifetime function as follows:
\begin{equation}\label{eq:tauNet}
\tau(\vec{k})
=  \frac{\TimeScale\,|\bma|^{\nu-\frac{2}{3}}}{(1+|\bma|^2)^{\nu/2}},
\qquad
\bma=\bma(\vec{k}),
\end{equation}
where $\TimeScale$ is characteristic timescale, $\nu$ is a tunable exponent,
and $\bma$ is a intermediary variable, which we refer to as the ``augmented'' wavevector.
We will define $\bma(\vec{k})$ in terms of a neural network $\bma = \mathcal{O}(k)$ in both limits $k\to 0$ and $k\to\infty$.
Therefore, the form of the fractional exponents in~\eqref{eq:tauNet} will allow us to control the asymptotic behavior of $\tau(\vec{k})$.
This is important since the asymptotic slopes are objects of discussion in the literature.
Thus, while the slope at zero is controlled by $\nu$, the well-known slope at infinity,\cite{landau1987fluid,mann1994spatial,pope2001turbulent} $k^{-2/3}$, is recovered for any value of~$\nu$.

We define
\begin{equation}
\label{eq:augmented_wavenumber}
	\bma(\vec{k}) := \mathrm{abs}(\vec{k}) + \mathrm{NN}(\mathrm{abs}(\vec{k})).
\end{equation}
Here, $\mathrm{abs}(\cdot)$ denotes the element-wise absolute value and $\mathrm{NN}(\cdot)$ is a fully-connected multilayer perceptron, namely
\begin{equation}\label{eq:NN}
	\mathrm{NN}(\bmy)
	:=
	W_{n}\, \mathrm{ReLU} \circ \cdots \circ \mathrm{ReLU} \circ W_1 \bmy
			,
\end{equation}
where $\mathrm{ReLU}(\cdot)$ stands for the rectified linear activation function and $W_j$, $j=1,\ldots,n$, are dense matrices (no bias terms).
The choice of a ReLU function is motivated by the requirement that $\mathrm{NN}(\vec{0})=\vec{0}$, thus, $\bma(\vec{0}) = \vec{0}$.
Other activation functions, such as the softplus activation function will not guarantee this property.

Note that taking the absolute value of the first argument in~\cref{eq:augmented_wavenumber} provides a reflection symmetry in~$\tau(\vec{k})$ with respect to each Cartesian axis.
This is obviously more than necessary to satisfy symmetry in the $k_2$-component.
Our experience has shown that that there is no benefit to the accuracy of the model, when fitting one-point spectra data, if we relax the symmetry the $k_1$- and $k_3$-components.

We use one input and one output layer, each of size~$3$, accompanied by $n-2$ hidden layers of size~$m$.
In other words, $W_1\in\R^{3\times m}$, $W_j\in\R^{m\times m}$, $j=2,\ldots, n-1$, and $W_{n}\in\R^{m\times 3}$, whose sum total of entries constitute a new vector of learnable parameters~$\vec{\theta}_{\mathrm{NN}}$.
Note that if any $W_j = 0$, $j=1,\ldots n$, then $\mathrm{NN} = 0$.
This forces the magnitude of the augmented wavevector to agree with the true wavenumber, $|\bma| = k$.
From this point-of-view the neural network acts in a way like a piecewise-linear perturbation of the original wavevector.

Once the eddy lifetime function~\cref{eq:tauNet} has been substituted into the rapid distortion model $\Phi(\bmk,\tau)$, we arrive at the DRD spectral tensor model:
\begin{equation}
	\Phi^{\rm DRD}(\bmk,\vec{\theta})
	:=
	\Phi(\bmk,\tau(\vec{k}))
	,
\end{equation}
where the vector of all learnable parameters, i.e.,
\begin{equation}\label{eq:params}
	\vec{\theta} = \{ \Magnitude, \LengthScale, \TimeScale, \nu, \vec{\theta}_{\mathrm{NN}} \}, \end{equation}
consists of the spectrum amplitude~$\Magnitude := c_0^2\, \varepsilon^{2/3}/u_\ast^2$, the characteristic length and time sales, $\LengthScale$ and $\TimeScale$, respectively, the exponent~$\nu$ and the weights~$\vec{\theta}_{\mathrm{NN}}$ of the neural network~\eqref{eq:NN}.

\section{Calibration}
\label{sec:Calibration}

In this section, we discuss various aspects of the model regression problem which calibrates the eddy lifetime $\tau(\bmk)$ together with the other free parameters in the DRD model~\cref{eq:params}.
More specifically, we document the problem formulation, state important details from our implementation,\cite{PythonCode} and summarize our results.

\subsection{Optimization problem formulation}

Let us consider the one-point spectra~\cref{eq:Kaimal} on the interval~$\dom=[0.1, 100]$.
We wish to find the model parameters~$\vec{\theta}$ which best fit this data.
To set the stage, define
\begin{equation}\label{eq:OPS}
\tilde{F}_{ij}(k_1;\vec{\theta})
=
\int\limits_{-\infty}^\infty\int\limits_{-\infty}^\infty \Phi_{ij}^{\rm DRD}(\bmk,\vec{\theta}) \dd k_2 \dd k_3
,
\end{equation}
and
\begin{equation}
\tilde{J}_i(f;\vec{\theta}) = \Magnitude\,k_1\tilde{F}_{ii}(k_1z;\vec{\theta})
,
\quad
i=1,2,3
,
\end{equation}
and $\tilde{J}_4(f;\vec{\theta}) = \Magnitude\,k_1\tilde{F}_{13}(k_1z;\vec{\theta})$, where $f=(2\pi)^{-1}k_1 z$.
Note that the learnable magnitude~$\Magnitude$ involves the constants $c_0^2\, \varepsilon^{2/3}$ and $u_\ast^2$.

To find the optimal parameter vector~$\vec{\theta}_{\rm opt}$, we solve the optimization problem
\begin{equation}
\label{eq:ModelRegression}
\min_{\vec{\theta}}\big\{\loss[\vec{\theta}] + \alpha\pen[\vec{\theta}] + \beta\reg[\vec{\theta}_{\mathrm{NN}}]\big\},
\end{equation}
where~$\alpha,\beta\ge0$ and each function $\loss[\vec{\theta}]$, $\pen[\vec{\theta}]$, and $\reg[\vec{\theta}_{\mathrm{NN}}]$ is defined as follows.
The mean squared error is defined as
\begin{equation}\label{eq:loss}
\loss[\vec{\theta}] : = 
\frac{1}{L}
\sum_{i=1}^{4} 
\sum_{j=1}^{L}
\big(\log|\textstyle{J_i}(f_j)|- \log|\textstyle{\tilde{J}_i(f_j,\vec{\theta})}|\big)^2,
\end{equation}
where $L$ is the number of the data points $f_j\in\mathcal{D}$.
The data $J_i(f_j)$ is evaluated using the Kaimal spectra~\eqref{eq:KaimalSu}--\eqref{eq:KaimalCross}.
The penalization term is defined as
\begin{equation}\label{eq:Penalty}
\pen[\vec{\theta}] : = 
\frac{1}{\abs{\dom}}
\sum_{i=1}^{4}
\norm[\mathcal{D}]{\mathrm{ReLU}\!\left(\frac{\partial^2\log\abs{\tilde{J}_i(\cdot,\vec{\theta})}}{(\partial\,\log k_1)^2}\right)}^2
,
\end{equation}
where 
$\abs{\dom} = \norm[\mathcal{D}]{1}$ 
and the norm $\norm[\mathcal{D}]{\cdot}$ is defined
\begin{equation}
\label{eq:lognorm}
	\norm[\mathcal{D}]{g}^2 := \int_{\dom}\abs{g(f)}^2\d(\log f).	
\end{equation}
This term penalizes the curvature of the log-spectra in order to obtain convex curves and to minimize oscillations, thus protecting against overfitting.
Associating $\vec{\theta}_{\mathrm{NN}}$ with a vector $(\theta_1,\ldots,\theta_N)\in\mathbb{R}^N$, we write
\begin{equation}\label{eq:Regularization}
	\reg[\vec{\theta}_{\mathrm{NN}}] : =
		\frac{1}{N} \sum_{i=1}^N \theta_{\mathrm{NN},i}^2
	,
\end{equation}
where $N$ denotes the total number of weights of the neural network~\cref{eq:NN}.
The term accelerates convergence and also helps to avoid overfitting.

\subsection{Implementation details}
\label{sub:implementation_details}

\begin{figure}[!ht]
	\centering
	\includegraphics[width=\linewidth]{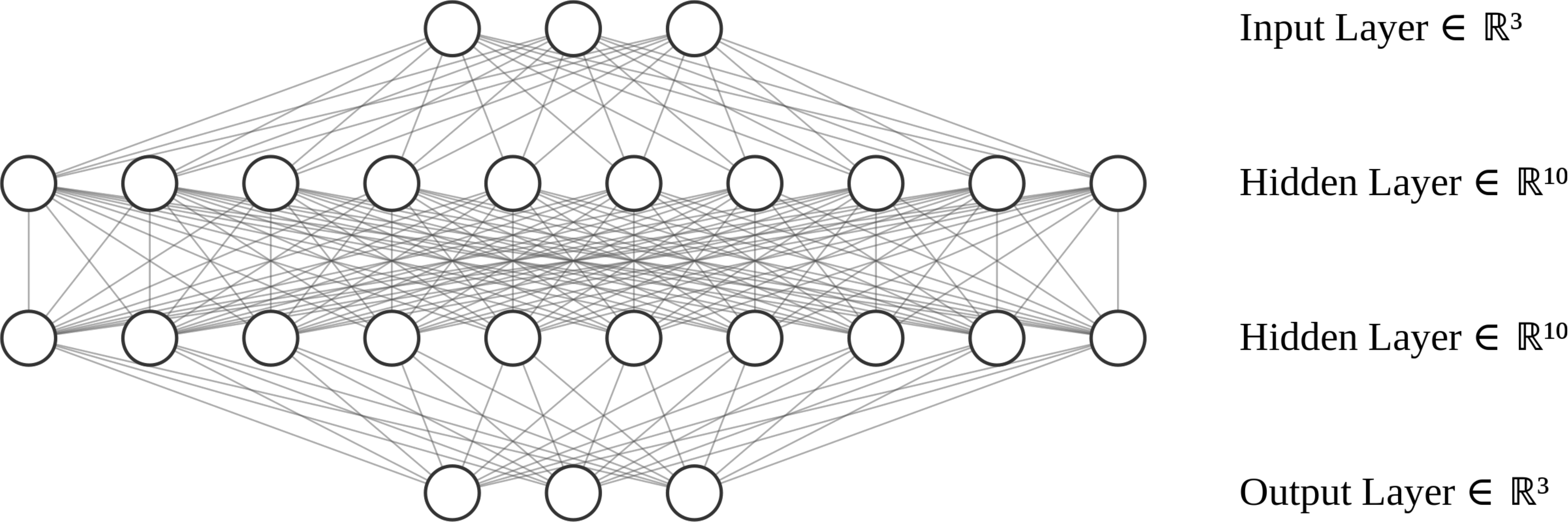}
	\caption{\label{fig:nn} Fully-connected neural network with $2$~hidden layers of size~$10$.
	Figure is created using NN-SVG tool.~\cite{lenail2019nn}}
\end{figure}

The model regression problem~\cref{eq:ModelRegression} is implemented and solved using the PyTorch package~\cite{NEURIPS2019_9015}.
In particular, we use a full-batch L-BFGS optimization algorithm with the strong Wolfe line search method and learning rate equal to~$1$.
We use $n=2$ hidden layers of size $m=10$ (see~\Cref{fig:nn}) and begin by using the penalty and regularization parameters $\alpha= 1$ and $\beta=10^{-5}$, respectively.

The integrals in~\eqref{eq:OPS} are truncated and approximated using the trapezoidal quadrature rule on a logarithmic grid from $10^{-3}$ to $10^3$ with $100\times100$ nodes spanning the frequencies $k_2$ and $k_3$.
The second derivative in~\eqref{eq:Penalty} is approximated with central finite differences.
And the norm~\cref{eq:lognorm} is approximated using the trapezoidal rule.

\subsection{Results and discussions}
\label{sub:results}

The IEC-recommended spectral tensor model~\cref{eq:IECTensor} is calibrated in Ref.~\onlinecite{mann1998wind} to fit the Kaimal spectra.~\cite{kaimal1972spectral,kaimal1994atmospheric}
In our notation, its three free parameters are $\LengthScale$, $\TimeScale$, and $\Magnitude$.
We use the fitted values from Ref.~\onlinecite{mann1998wind}, i.e., $\LengthScale=0.59$, $\TimeScale=3.9$, $\Magnitude=3.2$, to compare against the DRD model.

\begin{figure*}[!ht]
	\centering
	\begin{subfigure}[!ht]{0.49\linewidth}
		\includegraphics[width=\linewidth]{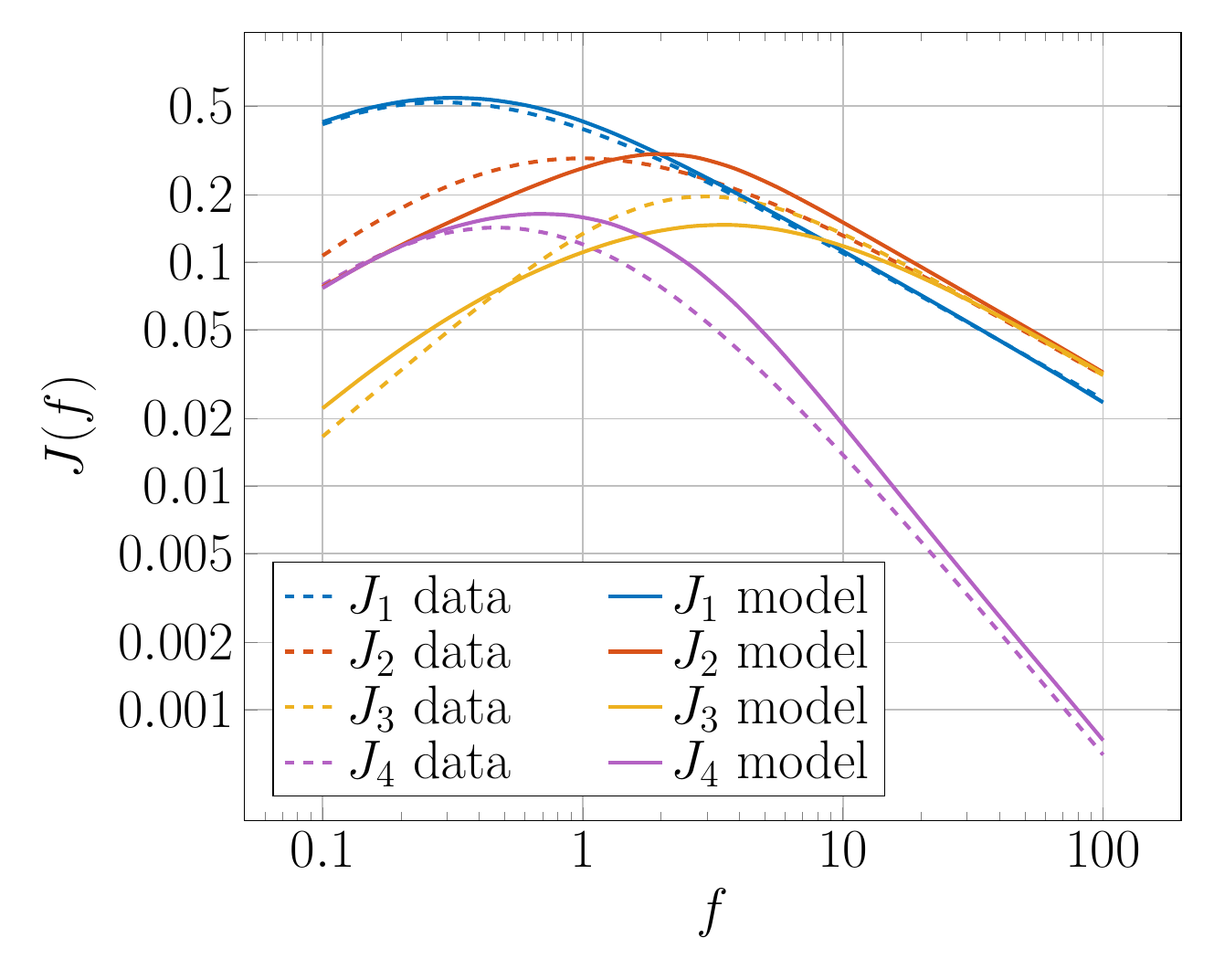}
		\caption{\label{fig:Mann_Kaimal} IEC 61400-1 model.}
	\end{subfigure}	
	\centering
	\hfill
	\begin{subfigure}[!ht]{0.49\linewidth}		
		\includegraphics[width=\linewidth]{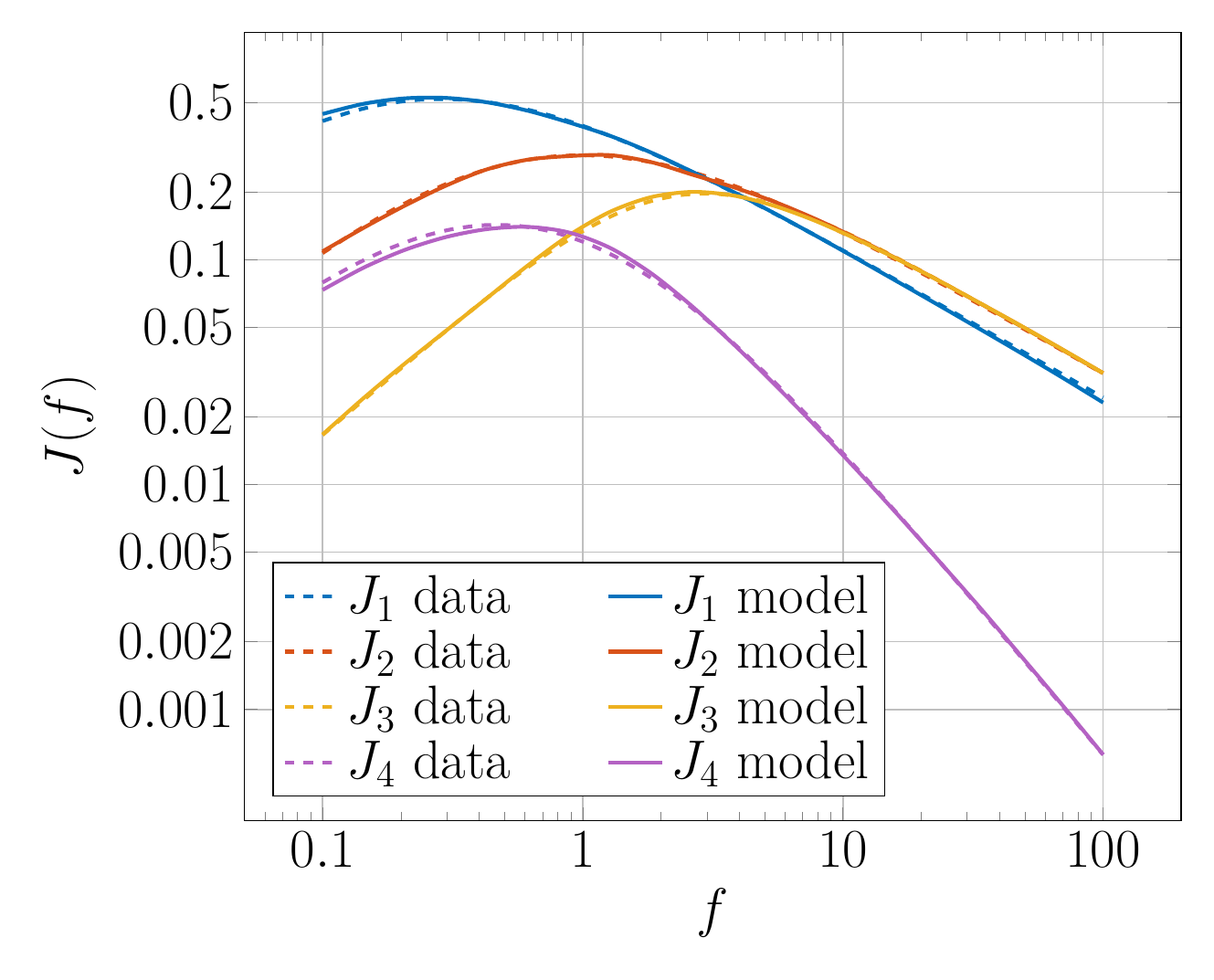}
		\caption{\label{fig:tauNet_Kaimal} DRD model with $\nu = -\frac{1}{3}$ (fixed).}
	\end{subfigure}
	\caption{\label{fig:Kaimal}Fits of the Kaimal spectra using two different spectral models.}
\end{figure*}

\begin{figure}[!ht]
	\centering	
	\includegraphics[width=\linewidth]{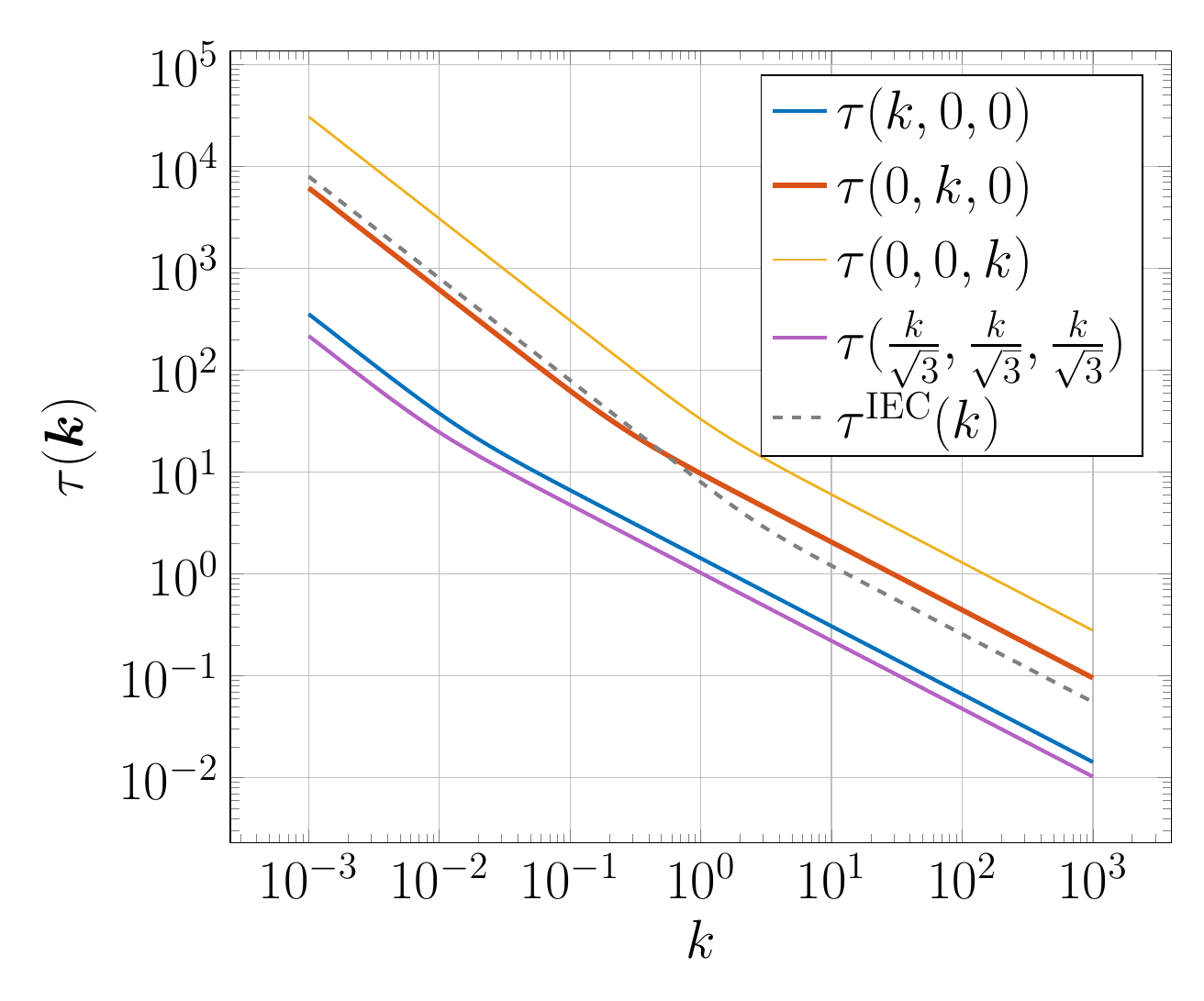}
	\caption{\label{fig:tau_plot}
			Cuts of the eddy lifetime function~$\tau(k_1,k_2,k_3)$ (solid lines) defined by~\eqref{eq:tauNet} with parameters corresponding to the fit in~\Cref{fig:tauNet_Kaimal}.
			The dashed line corresponds to the eddy lifetime function defined in~\eqref{eq:tau_Mann}, cf.~\Cref{fig:Mann_Kaimal}.}
\end{figure}

\begin{figure}[!ht]
	\centering	
		\includegraphics[width=\linewidth]{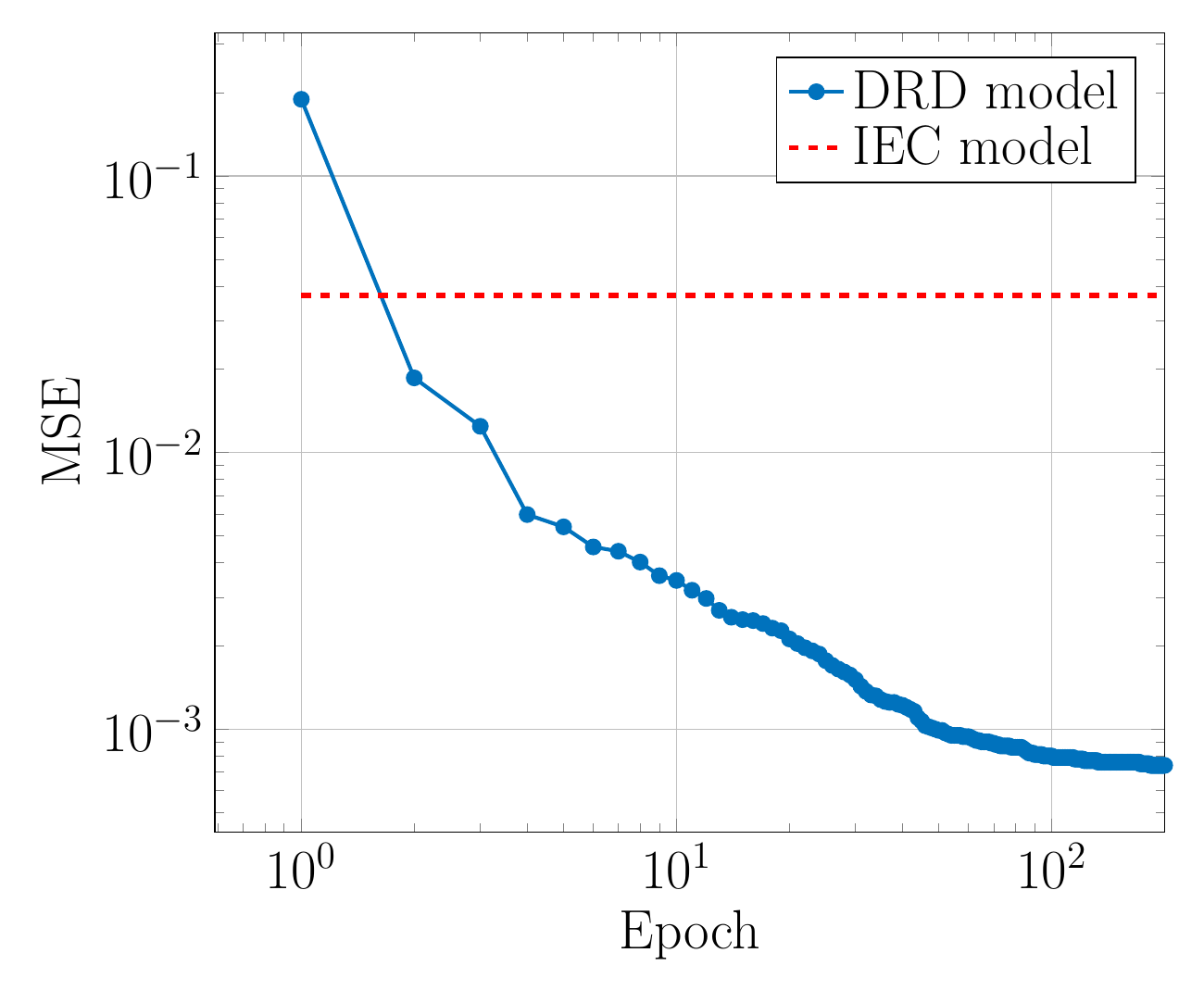}
	\caption{\label{fig:convergence} Convergence of the mean squared error~\eqref{eq:loss} during training to fit the one-point spectra in~\Cref{fig:tauNet_Kaimal}.
		The accuracy is more than an order of magnitude better than the IEC 61400-1 model, which achieves the fit shown in~\Cref{fig:Mann_Kaimal}.}
\end{figure}

In our first experiment, we \emph{fix} the exponent ${\nu=-\frac{1}{3}}$.
This is done so that $\tau(\bmk)$ matches the slope of $\tau^{\rm IEC}$ as $k\to 0$; in other words, so that
\begin{equation*}
	\tau(\bmk) \propto \tau^{\rm IEC}(k) \propto k^{-1}
	\quad
	\text{for }
	k\to 0
	.
\end{equation*}
We then set the initial scales of the DRD model to, $\LengthScale=0.59$, $\TimeScale=3.9$, $\Magnitude=3.2$ and initialize the weights $\vec{\theta}_{\mathrm{NN}}$ with random values drawn from additive Gaussian white noise with variance $\num{e-2}$.
We consider the range $\dom=[0.1, 100]$ and sample the data points $f_j\in\dom$ using a logarithmic grid of $N_{nodes}=20$~nodes.
The result of fitting the Kaimal spectra can be seen in~\Cref{fig:tauNet_Kaimal}.
This can be compared to the best fit produced by~\cref{eq:IECTensor}, depicted in~\Cref{fig:Mann_Kaimal}.
We observe that the DRD model presents a much better fit than the Mann uniform shear model, by a clear order of magnitude.
To show the learned eddy lifetime~$\tau(k_1,k_2,k_3)$, \Cref{fig:tau_plot} compares several cross-sections to the eddy lifetime curve given by the formula~\eqref{eq:tau_Mann}.
In~\Cref{fig:convergence}, we plot the convergence of the mean squared error~\eqref{eq:loss} throughout the optimization procedure.

In our second experiment, we proceed by fitting a perturbation of the Kaimal spectra with multiplicative log-normal noise.
Note that in this case, the use of the penalty term~\eqref{eq:Penalty} is important to avoid overfitting the noisy data.
On the other hand, the regularization term must be tuned to avoid local minima of the loss function~\cref{eq:ModelRegression}.
For best results, we increase the regularization parameter to $\beta=10^{-2}$.
All other parameters described in~\cref{sub:implementation_details} are left unchanged.

\begin{figure}[!ht]
	\centering	
		\includegraphics[width=\linewidth]{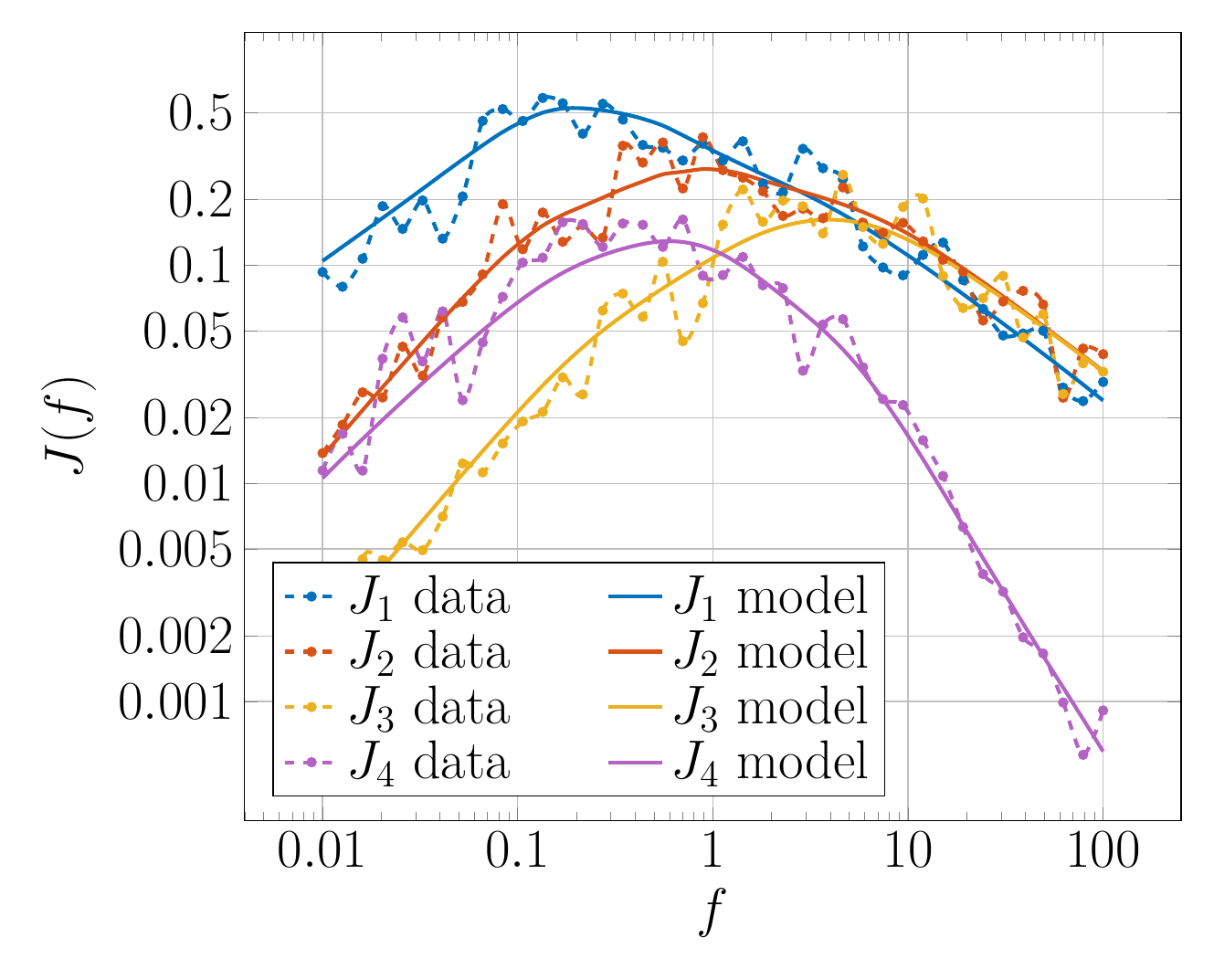}
	\caption{\label{fig:tauNet_Kaimal_noisy} Fitting a noisy Kaimal spectra with the DRD model.
	}
\end{figure}

In this example, we illustrate the possibility of learning the slope of the eddy lifetime function in the energy-containing subrange.
More specifically, we now suppose that the exponent~$\nu$ is a learnable parameter, and we aim to calibrate it in addition to the remaining model parameters.
To this end, we consider the extended range $\mathcal{D}=[0.01,100]$, discretized with a logarithmic grid of $N_{nodes}=40$ nodes.
It is necessary to extended the range of the the data into low wavenumbers, in order to get an accurate estimate of $\nu$.
Our results are shown in~\Cref{fig:tauNet_Kaimal_noisy}.
Here, we see a good fit to the data.
The calibrated slope is given by~$\nu=-0.55$.

\section{Random field generation} \label{sec:implementation_details}

In this section, we briefly discuss computational aspects of turbulence generation with the DRD and other spectral tensor models.
We take particular care to focus on aspects related to the use of such models for generating turbulent inlet conditions for LESs.

\subsection{Turbulent inlet conditions} \label{sub:projection}

Synthetic turbulent inlet conditions are used to drive many important turbulence simulations.\cite{tabor2010inlet,wu2017inflow}
For numerical wind tunnel simulations, it is important that the statistics of the input wind match the simulated environment.
This can be accomplished by relying on a synthetic turbulence model,\cite{sanderse2011review} such as the DRD introduced above.

In this section, we choose to follow an approach established in the wind engineering industry; see, e.g., Refs.~\onlinecite{Michalski2011,MICHALSKI201596,Andre2015} and references therein.
In this approach, a long channel of synthetic turbulence is treated as an inflowing velocity field with updated cross-sections projected onto the LES domain boundary at each time step.
For a 3D depiction of the process, including the accompanied LES, we refer to Ref.~\onlinecite[Figure~7]{keith2020fractional}.

An established technique to accomplish this process is to create a long unbroken section of synthetic wind offline and read in a moving cross-section of it at each time step of the LES simulation.
In this next subsection, an alternative on-the-fly domain decomposition technique is described.
This technique was used to generate the snapshots depicted in~\cref{fig:Snapshots} as well as the LES simulation in Ref.~\onlinecite[Figure~7]{keith2020fractional}.
Before we begin, we note that there is nothing specific to the DRD model about this technique.
It may be applied to any spectral turbulence model, including the model in IEC 61400-1.\cite{mann1998wind,tc882005iec}

\subsection{On-the-fly generation} \label{sub:on_the_fly_generation}

The representation of the turbulent fluctuations~\eqref{eq:FourierStieltjes} can be formally written as a convolution of a covariance kernel with Gaussian noise~$\bxi$ in the physical domain:
\begin{equation}\label{eq:Correlate}
\bfu
=
\mathcal{F}^{-1}\mathcal{G}\hat{\bxi}
=
\mathcal{F}^{-1}\mathcal{G}\mathcal{F}\bxi,
\end{equation}
where $\mathcal{F}$ stands for the Fourier transform and the operator~$\mathcal{G}$ corresponds to point-wise multiplication by $\bmG(\bmk)$, which is any positive-definite ``square root'' of the spectral tensor, satisfying $\bmG(\bmk)\, \bmG^\ast(\bmk) = \Phi(\bmk)$.

Experience shows that solving all-at-once for a unbroken section of synthetic wind, long enough to be fed into a complete CFD simulation, can be very costly.
First of all, the cost of solving~\cref{eq:Correlate} scales at best log-linearly with the domain length.
Second of all, storing the solution data may take up a very large amount of computer resources.
In fact, for a standard 600s LES simulation, the entire synthetic wind field may require more than 10GB to store.

It turns out the much of this computational cost can be avoided simply by piecing together the random field box-by-box.
This technique can be combined with inlet condition generation; see, e.g., Ref.~\onlinecite[Figure~7]{keith2020fractional}.
In this application, the technique allows for on-the-fly synthetic wind generation, since only a small box of wind pertaining to the given time step needs to be generated in order to progress the simulation.

\begin{figure*}[!ht]
\centering
	\begin{minipage}{0.8\linewidth}
	\centering
		\includegraphics[clip=true, trim = 3cm 1cm 3.1cm 1.5cm, width=\textwidth]{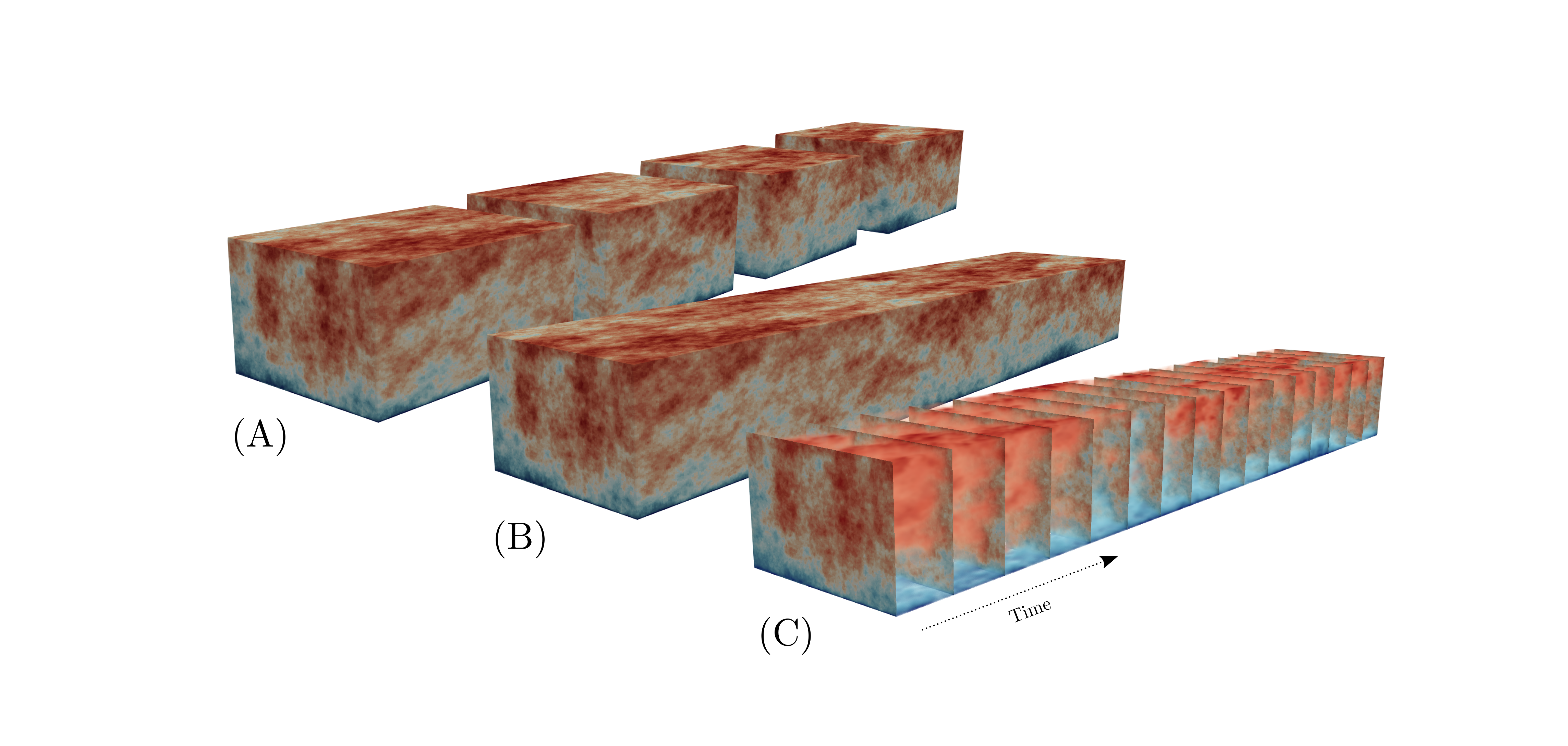}		
			\end{minipage}
	\begin{minipage}{0.075\linewidth}
	\centering
		\includegraphics[width=\textwidth]{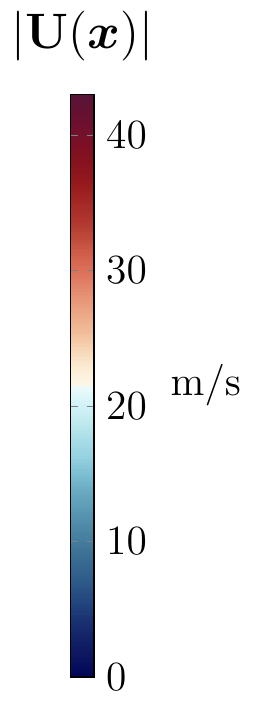}
	\end{minipage}
		\caption{\label{fig:Snapshots}
	Turbulent fluctuations generated with the DRD model with learned parameters from fitting the Kaimal spectrum (cf. \Cref{fig:tauNet_Kaimal}) and superimposed on the mean profile~\cref{eq:MeanProfileLog}:
	(A) wind generated box-by-box; (B) combined boxes: (C) snapshots. }
\end{figure*}

\begin{figure}
\centering
			\includegraphics[width=0.95\linewidth]{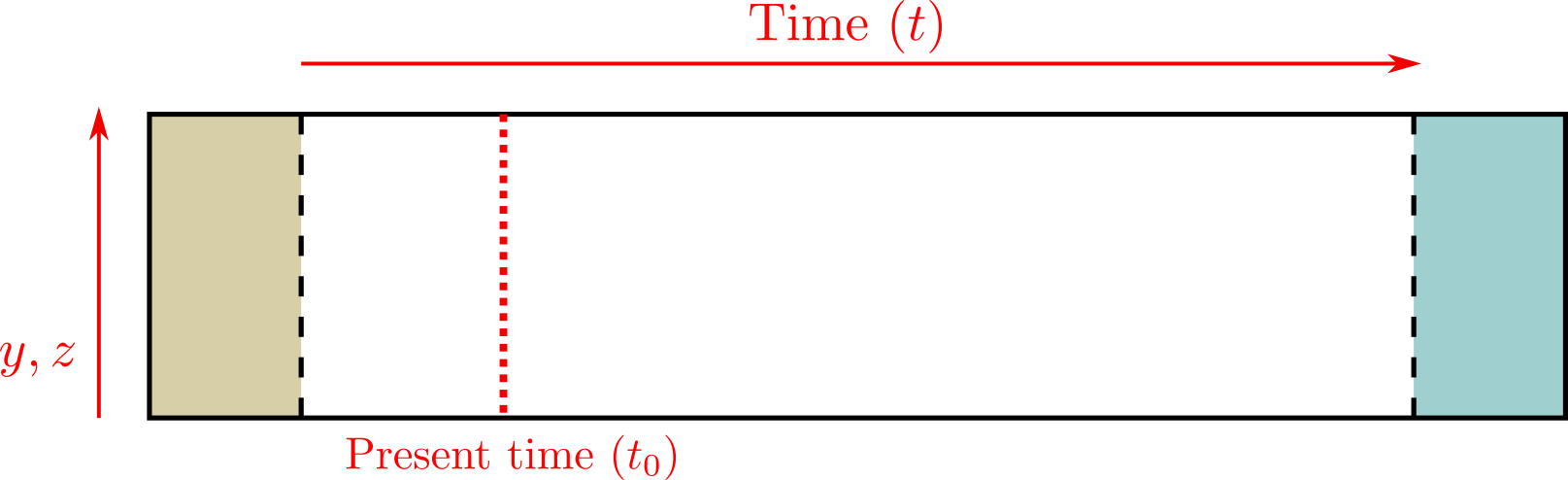}
	\caption{\label{fig:Buffer} Simplified diagram of a 2D computational domain with buffer regions (colored) at only two opposing ends.
	The fluctuations in the buffer regions are highly correlated so they are typically discarded during post-processing.
	}
\end{figure}

Begin with any box domain and extend it on all sides by a ``buffer region.''
Seeding a Cartesian grid inside the extended box with additive white Gaussian noise and then computing the convolution~\cref{eq:Correlate} will results in a periodic random vector field.
Restricting the solution to the original domain will break this periodicity.
Indeed, it will result in a random field with a (non-zero) correlation between opposing ends of the original domain.
This correlation is controlled by the size of the buffer region and as well as the shape of the covariance kernel.\cite{Khristenko2019}
It is common practice to extend the buffer region until this correlation is negligible. 
A simplified diagram which shows a long 2D box (this can be thought of as a transversal cross-section of a long 3D turbulence box) with a buffer only at two opposing end is depicted in~\cref{fig:Buffer}.

\begin{figure}[!ht]
	\centering
	\includegraphics[width=0.95\linewidth]{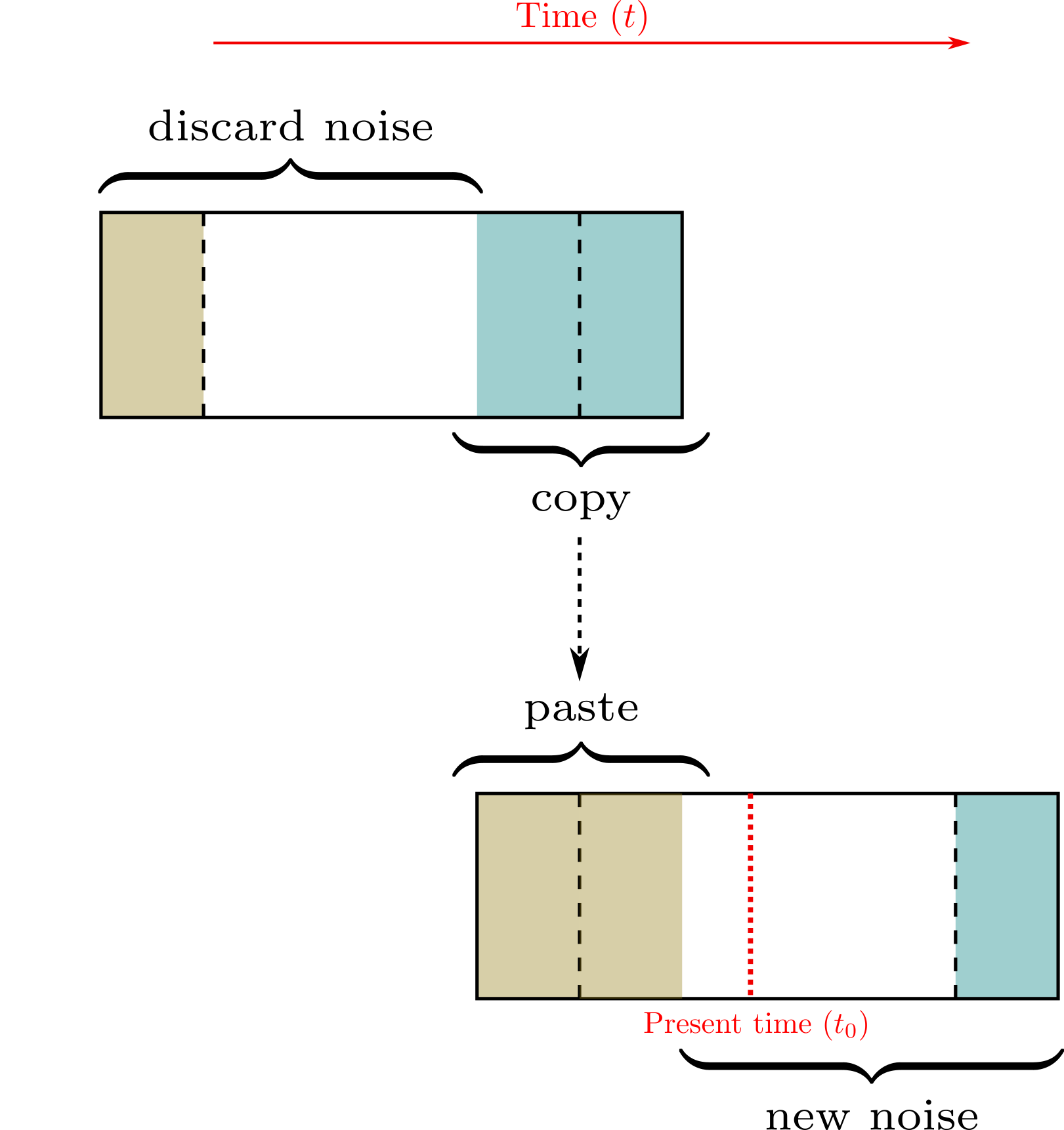}
	\caption{\label{fig:OnTheFly}Technique to generate a continuous wind field box-by-box.
	Note that \textit{only} the noise is being copied from the end of one block to the beginning of the next block.
	The turbulence fluctuations $\mathbf{u}$ are still recomputed block-by-block, using the partially shared noise.
	}
\end{figure}

In order to generate matching wind in a neighboring box, it is required for the interface of the new neighboring buffer region (cf. the ``copy-paste'' lines in \cref{fig:OnTheFly}) to align with the interface of the old neighboring buffer region and that common Gaussian noise be used in the overlapping domain.
After seeding the remainder of the new box with new additive white Gaussian noise and applying~\cref{eq:Correlate}, one arrives at a new box of wind matching its neighbor, at their common interfaces, up to an accuracy again controlled by the size of the buffer region.
Once the solutions in the buffer regions are discarded, the original neighboring boxes of wind field can be grouped together to form a large unbroken wind field or processed into snapshots and fed into a CFD simulation; cf. \cref{fig:Snapshots}~(C).

To understand why this technique works, note that the common noise in the overlap forces the two wind fields to match at their buffer region interfaces, at least up to the accuracy allowed by the size of the buffers.
Moving away from the interface, the wind field changes at a fixed regularity, with the influence of the common overlap steadily diminishing.\cite{Khristenko2019}
This strategy can be applied to any homogeneous spectral tensor turbulence model, not only the DRD model or the Mann uniform shear model.

\section{Conclusion} \label{sec:conclusion}

We have presented the deep rapid distortion (DRD) model: a new nonlocal, data-driven spectral model for the atmospheric boundary layer.
The DRD model is derived from rapid distortion theory and leverages an eddy lifetime function parameterized in terms of a carefully-chosen neural network.
The model is calibrated using a regression problem involving the fitting of (possibly noisy) one-point spectrum data.
Using this data, we witness exceptional accuracy with the DRD model, especially when compared to the current IEC standard.
Finally, we present a domain decomposition method for using the DRD model and other spectral models to generate synthetic turbulence.
For the purpose of reproduction and wider adoption, this work is accompanied by an open-source Python implementation.\cite{PythonCode}

\begin{acknowledgments}
We wish to thank Georgios Deskos for helpful discussions and comments on the manuscript.
We also wish to thank Jakob Mann for providing us with reference values of the IEC 61400-1 model, which helped us debug our own implementation.

This project has received funding from the European Union's Horizon 2020 research and innovation programme under grant agreement No 800898.
This work was also partly supported by the German Research Foundation by grant WO671/11-1.

This work was performed under the auspices of the U.S. Department of Energy by Lawrence Livermore National Laboratory under Contract DE-AC52-07NA27344, LLNL-JRNL-824304-DRAFT. This document was prepared as an account of work sponsored by an agency of the United States government. Neither the United States government nor Lawrence Livermore National Security, LLC, nor any of their employees makes any warranty, expressed or implied, or assumes any legal liability or responsibility for the accuracy, completeness, or usefulness of any information, apparatus, product, or process disclosed, or represents that its use would not infringe privately owned rights. Reference herein to any specific commercial product, process, or service by trade name, trademark, manufacturer, or otherwise does not necessarily constitute or imply its endorsement, recommendation, or favoring by the United States government or Lawrence Livermore National Security, LLC. The views and opinions of authors expressed herein do not necessarily state or reflect those of the United States government or Lawrence Livermore National Security, LLC, and shall not be used for advertising or product endorsement purposes.
\end{acknowledgments}

\section*{Data Availability}
The data that supports the findings of this study are available within the article and its supplementary material.\cite{PythonCode}

\phantomsection\bibliography{main}

\end{document}